\documentclass[10pt,amsmath,amssymb,aps,prl,twocolumn,superscriptaddress]{revtex4}
\usepackage{graphicx}

\DeclareMathOperator{\re}{Re}
\DeclareMathOperator{\im}{Im}

\begin{document}
\title{Aharonov-Casher Effect for Plasmons in a Ring of Josephson Junctions} 
\author{Roman S\"usstrunk}
\affiliation{Institute for Theoretical Physics, ETH Zurich, 8093 Zurich, Switzerland}
\affiliation{Department of Physics, Yale University, New Haven, CT 06520, USA}
\author{Ion Garate}
\affiliation{Department of Physics, Yale University, New Haven, CT 06520, USA}
\affiliation{D\'epartement de Physique, Universit\'e de Sherbrooke, Sherbrooke, Qu\'ebec, Canada J1K 2R1}
\author{Leonid I.\ Glazman}
\affiliation{Department of Physics, Yale University, New Haven, CT 06520, USA}
\date{\today}

\begin{abstract}
Phase slips in a one-dimensional closed array of Josephson junctions hybridize the persistent current states (PCS) and plasmon branches of excitations. The interference between phase slips passing through different junctions of the array makes the hybridization sensitive to the charges of the superconducting islands comprising the array. This in turn results in the Aharonov-Casher effect for plasmons, which in absence of phase slips are insensitive to island charges.
\end{abstract}
\maketitle

For over two decades, arrays of small superconducting islands connected by Josephson junctions have been a popular platform for experimental studies of quantum fluctuations in superconductors ~\cite{fazio:01a}. 
This popularity has originated in part from the tunability of the islands' charging energy and the junctions' Josephson energy, which has enabled the design of arrays with a desired level of quantum fluctuations of the phases of the order parameter.

The most fundamental manifestation of the quantum fluctuations of phase is a quantum phase slip. 
The proliferation of phase slips in long nanowires or Josephson junction arrays is predicted to lead to a quantum phase transition between superconducting and insulating states~\cite{bradley:84a,sondhi:97a}. 
In addition, coherent phase slips interfere with each other, giving rise to the Aharonov-Casher effect~\cite{ac:84}. 
An unambiguous observation of these predictions in dc measurements has proven to be difficult~\cite{fazio:01a,chow:98a,pop:12}. 
However, an evidence of the Aharonov-Casher effect in long arrays has recently been seen in spectroscopic measurements~\cite{manucharyan:12a} performed on a ring of junctions pierced by a magnetic flux. 
The measurements~\cite{manucharyan:12a,manucharyan:09a} focused on the avoided crossing of the two lowest-energy states. 
These two states carry counter-propagating persistent currents and become degenerate if the magnetic flux equals to a half-integer of flux quantum; phase slips remove the 
degeneracy. 
The Aharonov-Casher effect reveals itself through the modulation of the avoided-crossing gap by charges in the superconducting islands. 
To interpret the experiment of Ref.~\cite{manucharyan:12a}, it was sufficient to account for the hybridization of two otherwise degenerate lowest-energy many-body states~\cite{manucharyan:12a,ivanov:01a,matveev:02a}. 
However, measurements of the same type~\cite{manucharyan:09a,manucharyan:12a,masluk:12a,bell:12a} routinely display a rich, flux-dependent level structure at higher energies. 
This structure is poorly understood and, to our knowledge, the effect of phase slips on the spectrum of excitations has not been studied theoretically.

\begin{figure}[tb]
\centering 
\includegraphics{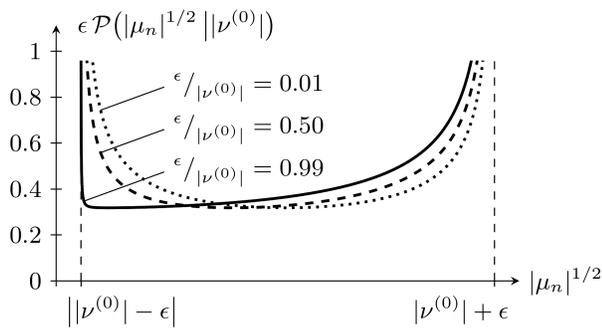}
\caption{Conditional distribution of splittings of the one-plasmon level for given $|\nu^{(0)}|$, according to Eq.~(\ref{eq:conditionPDFofMu}). 
Shown is the case $\epsilon\leq |\nu^{(0)}|$.}
 \label{fig:probDensFctOnePlasmon}
\end{figure}

In this work, we investigate the effect of quantum phase slips on the spectrum of plasmons in a one-dimensional array of Josephson junctions forming a ring. 
In a dense plasmon spectrum, the structure of the multiple avoided crossings and their sensitivity to the charges (Aharonov-Casher effect) turns out to be quite different from the single avoided crossing~\cite{ivanov:01a,matveev:02a} occurring in the ground state, see Figs.~\ref{fig:probDensFctOnePlasmon} and \ref{fig:spectrum}.

\begin{figure}[htb]
 \centering
 \includegraphics{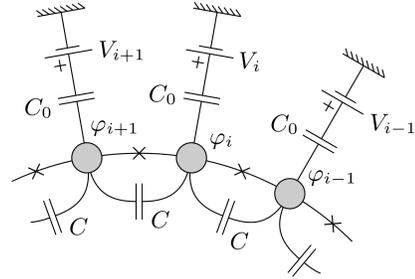}
 \caption{Segment of a ring of Josephson junctions. Grey circles represent superconducting islands, which are connected by ideal Josephson junctions (crosses).}
 \label{fig:model}
\end{figure}

\begin{figure}[htb]
 \centering
 \includegraphics{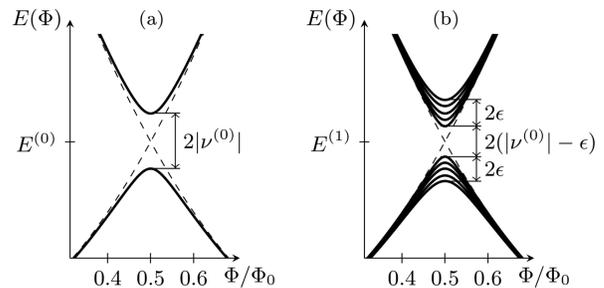}
 \caption{Schematic energy spectrum for a ring of $N$ Josephson junctions (Fig.~\ref{fig:model}) with Josephson energy $E_J$, small ground capacitance $C_0$, and a given realization of random gate-induced charges.
The ring is pierced by a magnetic flux $\Phi$. 
(a) Spectrum in absence of plasmon excitations, where $E^{(0)}\equiv \pi^2 E_J/(2 N)+(N-1) \hbar\omega_p/2$ and $\omega_p$ is the plasmon frequency. 
(b) Spectrum in presence of one plasmon, where $E^{(1)}\equiv E^{(0)}+\hbar\omega_p$.
In this figure $|\nu^{(0)}|>\epsilon$. 
The non-degeneracy of the $2(N-1)$ one-plasmon energy levels is due to the Aharonov-Casher effect.
The dashed lines illustrate the energy spectrum in absence of phase slips.}
 \label{fig:spectrum}
\end{figure}

We consider a ring of $N\gg 1$ identical junctions (see Fig.~\ref{fig:model}) of Josephson energy $E_J$.  
The phase $\varphi_i$ of the order parameter may be taken~\cite{fazio:01a} uniform within the grain $i$. 
Each grain is capacitively coupled to a voltage source $V_i$ with gate capacitance $C_0$; this models inevitable~\cite{manucharyan:12a} quasi-static charges induced by the environment.  
Two neighboring grains are connected through a Josephson junction shunted by a capacitance $C$. 
The ring is threaded by a magnetic flux $\Phi$ that induces a persistent current \cite{matveev:02a}. 
Since 
the self-inductance of the ring is negligible, $\Phi$ is
close to the flux of the externally applied magnetic field. 

At $V_i=0$, the Hamiltonian of the ring is
\begin{equation}\label{eq:hamiltonOperator}
 \mathcal{H}=2e^2\sum_{i,j=1}^{N}\hat Q_i(C^{-1})_{ij}\hat Q_j+E_J\sum_{i=1}^{N}\left[1-\cos\left(\hat\gamma_i\right)\right],
\end{equation}
where the first and second terms are the kinetic and potential energies describing the energy cost of electrostatic charging and persistent currents, respectively.
$C_{ij}=\delta_{i,j}(2C+C_0)-(\delta_{i,j+1}+\delta_{i,j-1})\,C$ is the capacitance matrix, and $\hat\gamma_i=\hat\varphi_{i+1}-\hat\varphi_i+\delta/N$ are the gauge-invariant phase differences across the junctions, with $\delta\equiv 2\pi \Phi/\Phi_0$. 
The operator $\hat Q_i$ corresponds to the charge on island $i$ in units of $-2e$ and is canonically conjugate to $\hat\varphi_i$.
The topology of the ring enforces the constraint~\cite{tinkham:96a}
\begin{equation}\label{eq:ringCondition}
 \sum_{i=1}^{N}\hat\gamma_i -\delta = 0 \pmod {2\pi}.
\end{equation}
Equation~(\ref{eq:hamiltonOperator}) is supplemented by the boundary condition $ \psi(\varphi_1,\dots,\varphi_i+2\pi,\dots,\varphi_{N})=\psi(\varphi_1,\dots,\varphi_i,\dots,\varphi_{N})$, for all $i\in\{1,\dots,N\}$.
Because only phase differences enter Eq.~(\ref{eq:hamiltonOperator}), the problem is invariant under a simultaneous shift of all $\varphi_i$; the associated conserved quantity is the total charge $Q=\sum_i Q_i\in \mathbb{Z}$. 
Hereafter we restrict the discussion to a fixed value of $Q$.

When $V_i\neq 0$, induced charges $q_i=C_0V_i/2e$ break the rotational symmetry of the array and result in a shift of the charge operators in Eq.~(\ref{eq:hamiltonOperator}): $\hat Q_i\to \hat Q_i-q_i$. 
A unitary transformation eliminates the $q_i$-dependence from ${\cal H}$ and makes the wave functions independent of the total phase $\phi=\sum_i \varphi_i$. 
The outcome of this transformation is a Hamiltonian given by Eq.~(\ref{eq:hamiltonOperator}) (modulo constant terms), and a boundary condition~\cite{ivanov:01a}
\begin{multline}\label{eq:modifiedBoundaryCondition}
 \psi(\varphi_1\dots,\varphi_j+2\pi,\dots,\varphi_{N}) \\
  =e^{i2\pi\left(q_j-\frac{1}{N}\sum_{k=1}^N q_k-\frac{1}{N}Q\right)}\psi(\varphi_1,\dots,\varphi_j,\dots,\varphi_{N}).
 \end{multline}

The main objective of this work is to evaluate the energy spectrum for Eq.~(\ref{eq:hamiltonOperator}).
We are interested in the regime of $E_J$ dominating over the charging energy. To zeroth order in the charging energy, the phase variables become classical. The configurations minimizing the potential energy in Eq.~(\ref{eq:hamiltonOperator}) under the constraint of Eq.~(\ref{eq:ringCondition}) are characterized by an integer $m$, 
\begin{equation}\label{eq:currentStates}
\gamma_i\bmod 2\pi=(2\pi m+\delta)/N \,,
\end{equation}
 for all $i\in\{1,\dots,N\}$. 
These states, which host an electrical current whenever $\Phi\neq 0\pmod{\Phi_0}$, will be referred to as ``persistent current states'' (PCS) $|m\rangle$.
The specific value of $m$
leading to $\gamma_i\bmod 2\pi \in[-\pi/N,\pi/N]$ describes the classical ground state, while other values of $m$ denote the local minima of the potential and correspond to excited (metastable) states with higher currents. 

A small but nonzero charging energy results in quantum fluctuations of $\gamma_i$ around the (local) minima. 
For low-current states ($\gamma_i\bmod 2\pi \sim 1/N\ll 1$), small fluctuations are described by quadratic expansion~\cite{rastelli:12a},
\begin{equation}
\label{eq:unperturbedHamiltonian}
 \mathcal{H}_0=\sum_m E_m|m\rangle\langle m|+\sum_{l=1}^{N-1}\hbar\Omega_l\left(\frac{1}{2}+\hat{a}_l^\dagger \hat{a}_l\right)\,,
\end{equation}
of Eq.~(\ref{eq:hamiltonOperator}). Here $E_m\equiv (2\pi m+\delta)^2 E_J/(2 N)$,
and $\hat{a}_l^\dagger$ is the creation operator for a harmonic mode (``plasmon'') of frequency $\Omega_l$.
Note that only $N-1$ modes appear in $\mathcal{H}_0$~\cite{N}. 
Herein we restrict the discussion to zero- and one-plasmon states only, for which anharmonic effects~\cite{masluk:12a} are small.
In addition, we will be interested in low-current-carrying states for which the plasmon frequencies are essentially independent of the magnetic flux:
\begin{equation}\label{eq:frequenciesPlasmons}
 \Omega_l^2 =\omega_p^2\frac{1-\cos\left(\frac{2\pi}{N}l\right)}{1-\cos\left(\frac{2\pi}{N}l\right)+\frac{C_0}{2C}}\,,\qquad \omega_p^2=\frac{8E_JE_C}{\hbar^2}\,,
\end{equation}
where $\omega_p$ and $E_C=e^2/2C$ are the single-junction plasma frequency and charging energy, respectively.
To simplify notation, we have labeled the plasmon modes by $l\in\{1,\dots,N-1\}$ instead symmetric about $l=0$ wave numbers ~\cite{symmetric}. 
At $C_0\neq 0$, one-plasmon states are doubly degenerate (except for $l=N-l$) because $\Omega_l=\Omega_{N-l}$. At $C_0=0$, the degeneracy increases to $N-1$ because $\Omega_l=\omega_p$ for all $l$ (dispersionless plasmons).

The harmonic model $\mathcal{H}_0$ provides a good approximation for the energy spectrum of the ring, except in the vicinity of flux values for which two neighboring PCS ($|m\rangle$ and $|m-1\rangle$) become degenerate. 
These degeneracies, which emerge at $\Phi=\Phi_0/2 \pmod{\Phi_0}$, can be lifted by tunneling.
Tunneling between $|m\rangle$ and $|m-1\rangle$ involves a change of $\sim 2\pi$ in the phase difference across a junction, and is known as a quantum phase slip.
We neglect tunneling processes involving multiple simultaneous phase slips, because they are highly unlikely in our regime of interest ($E_J\gg E_C$).

The influence of phase-slips in the energy spectrum of ${\cal H}_0$ can be incorporated perturbatively by adding $\delta{\cal H}^{(0)}$ and $\delta{\cal H}^{(1)}$ in the zero- and one-plasmon subspaces, respectively, where
\begin{align}\label{eq:perturbation}
 \delta\mathcal{H}^{(0)} &=\sum_{m}\left(\nu^{(0)} |m-1\rangle\langle m|+ {\rm c.c.}\right)\nonumber\\
 \delta\mathcal{H}^{(1)} &=\sum_m \sum_{l,l'=1}^{N-1}\left(\nu^{(1)}_{l,l'}|m-1,l\rangle\langle l',m|+\text{c.c.}\right).
\end{align}
In Eq.~(\ref{eq:perturbation}), $\nu^{(0)}$ is the $|m\rangle\to|m-1\rangle$ tunneling amplitude in absence of plasmons, whereas $\nu_{l,l'}^{(1)}$ is the $|m\rangle\to|m-1\rangle$ tunneling amplitude in presence of one plasmon that gets scattered from mode $l'$ to $l$. 
For $\Phi\simeq\Phi_0/2 \pmod{\Phi_0}$ and $C_0\simeq 0$, the splitting between the two lowest-current levels is 
\begin{align}
  \label{eq:splitting}
 & \delta E^{(0)} =\left[(E_m-E_{m-1})^2+4 |\nu^{(0)}|^2\right]^{1/2} \,\,\,\nonumber\\
 & \delta E_n^{(1)} =\left[(E_m-E_{m-1})^2+4 |\mu_n|\right]^{1/2}\,
\end{align}
in the zero- and one-plasmon subspaces, respectively. 
In Eq.~(\ref{eq:splitting}), $\mu_n$ is the $n$-th eigenvalue of the matrix $\nu^{(1)\null\dagger}\nu^{(1)}$ ($n\in\{1,...,N-1\}$). 
The flux-dependence of the splitting comes mainly from $E_m(\Phi)$, while $\nu^{(0)}$ and $\mu_n$ can be evaluated at $\Phi=\Phi_0/2$.

Next, we proceed with the microscopic derivation of the tunneling amplitudes.
While the expression for $\nu^{(0)}$ is well-known, this Letter develops the first theory for $\nu^{(1)}$. 
For convenience we work with the non-gauge-invariant phase differences $\theta_i=\varphi_{i+1}-\varphi_i$. 
Without loss of generality, we consider the two lowest-current states at $\Phi=\Phi_0/2$ ($|m=0\rangle$ and $|m=-1\rangle$).
Each of them is represented in $\theta_i$-space by a manifold of points that minimize the potential energy [Eq.~(\ref{eq:currentStates})]. 
Representing $|m=0\rangle$ by $\{\theta_i\}=0$, its $N$ neighboring minima $\{\theta_i\}=-2\pi/N+\delta_{i,n}2\pi$ (where $n\in\{1,\dots,N\}$ all represent the $|m=-1\rangle$ PCS. 
Starting from $\{\theta_i\}=0$, each of the $|m=-1\rangle$ minima can be reached through an instanton on a least-action path. 
Along each of these paths a different junction (labeled by $n$)  undergoes a phase slip. The total $|m=0\rangle\to|m=-1\rangle$ tunneling amplitudes 
are the sums over the contributions of $N$ individual paths (denoted by $\nu_n^{(0)}$ and $\nu^{(1)}_{n;l,l'}$):
\begin{equation}\label{eq:tunnelingAmplitudes}
 \nu^{(0)}=\sum_{n=1}^N \nu_{n}^{(0)}\qquad\text{and}\qquad\nu^{(1)}_{l,l'}=\sum_{n=1}^N \nu_{n;l,l'}^{(1)} \,.
\end{equation}
Plasmons enhance the tunneling amplitudes via the oscillation energy available to assist a phase slip. 
Different junction variables oscillate with different amplitudes, depending on which of the $N-1$ plasmon modes is excited. 
That is why contributions of different junctions are different even as the instanton paths end in equivalent points.

The calculation of $\nu^{(1)}_{l,l'}$ requires specifying the plasmon wave functions. 
For $|m=0\rangle$ and $C_0\simeq 0$, they can be chosen as
\begin{equation}
 \psi_l=K_l\sum_{n=1}^{N-1}\theta_n e^{i\frac{2\pi ln}{N-1}} e^{-D^2\left(\sum_{n'=1}^{N-1}\theta_{n'} \right)^2}\prod_{j=1}^{N-1}e^{-D^2{\theta_j}^2},\nonumber
\end{equation}
where $l\in\{1,\dots,N-1\}$, $D^2\equiv\hbar\omega_p/(16E_C)$, and $K_l\propto [1+\delta_{l,N-1}(N^{1/2}-1)]$ is the normalization factor.
For the $N$ neighbouring minima representing $|m=-1\rangle$, we can pick exactly the same wave functions up to a shift in coordinate space, and denote them $\tilde{\psi}_l^{(n)}$. 
Due to Eq.~(\ref{eq:modifiedBoundaryCondition}), $\tilde{\psi}_l^{(n)}$ and $\tilde{\psi}_l^{(j\neq n)}$ are related by a phase factor.

The next step is to find the contribution of a single junction $n$, namely $\nu_{n;l,l'}^{(1)}$. 
Three difficulties become apparent.
First, a phase slip occurring in junction $n$ involves not only $\theta_n$ but  
also requires a shift of $-2\pi/N$ in all $\theta_{i\neq n}$. 
Second, a harmonic approximation of the potential energy of junction $n$ is no longer appropriate because $\gamma_n$ changes by almost $2\pi$. 
Third, having $N-1$ modes for $N$ junctions means that plasmons are delocalized along the ring, which in turn requires a prescription to determine how a plasmon assists the phase slip in junction $n$. 
We resolve the first two difficulties by (a) changing variables by a linear transformation $\{\theta_i\}\mapsto\{x_i\}$ ~\cite{supmat}  such that only $x_1$ changes during a phase slip, and (b) keeping the cosine term for the $n$-th junction while making a harmonic approximation for the rest.
The minimum $\{\theta_i\}=0$ corresponds to $\{x_i\}=0$ and $\{\theta_i\}=-2\pi/N+\delta_{i,n}2\pi$ becomes $\{x_i\}=\delta_{i,1}2\pi(N-1)/N$. 
Importantly, the proper choice of $\{x_i\}$ simplifies Eq.~(\ref{eq:hamiltonOperator}) into $\mathcal{H_{\text{eff}}}+
\mathcal{H}_{\text{harm}}$, where
\begin{align}
\label{eq:effectiveHamiltonian}
 \mathcal{H}_{\text{eff}} &=E_J\hspace{-1pt}\left[1-\cos\left(x_1+\frac{\delta}{N}\right)-\frac{\delta}{N}x_1+\frac{1}{2(N-1)}x_1^{\ 2}\right] \nonumber\\ 
  &+\frac{4E_C}{\hbar^2}\frac{N-1}{N}p_{x_1}^{\ 2}
\end{align}
is an effective Hamiltonian for junction $n$, and 
\begin{equation}
 \mathcal{H}_{\text{harm}}=\sum_{k=2}^{N-1}\hbar\omega_p \left(\frac{1}{2}+\hat{b}_k^\dagger \hat{b}_k\right)+\frac{E_J}{2N}\frac{N-1}{N}\pi^2\,
\end{equation}
 describes $N-2$ harmonic modes that are decoupled from $x_1$.
The operator $p_{x_1}$ in Eq.~(\ref{eq:effectiveHamiltonian}) is conjugate to $x_1$, and the linear and quadratic terms in $x_1$ arise from Eq.~(\ref{eq:hamiltonOperator}) and Eq.~(\ref{eq:ringCondition}). 

A harmonic approximation of $\mathcal{H}_{\text{eff}}+\mathcal{H}_{\text{harm}}$ about a minimum results in a set of modes $\{\hat{b}_k\}$, which are generally different from $\{\hat{a}_k\}$ of Eq.~(\ref{eq:unperturbedHamiltonian}).
In terms of these new modes, only $\hat{b}_1^\dagger$, associated with variable $x_1$, creates an excitation affecting tunneling in junction $n$: in absence of this plasmon the amplitude has some absolute value $\nu_0$, and if there is a plasmon the amplitude has a different absolute value $\nu_1$ 
~\cite{nu0}. 
Thus, the initially multidimensional tunneling problem is now reduced to a textbook one-dimensional problem.
Next, we expand $\psi_{l'}$ and $\tilde{\psi}_{l}^{(n)}$ in terms of the two sets of plasmon wave functions that follow from the harmonic approximation of $\mathcal{H}_{\text{eff}}+\mathcal{H}_{\text{harm}}$ about $x_1=0$ and $x_1=2\pi(N-1)/N$, respectively. 
This resolves the third difficulty mentioned above. 
From the expansion, we arrive at
\begin{equation}\label{eq:nu1}
 \begin{aligned}
 \nu_{n<N;l,l'}^{(1)} &= z_n \nu_0\delta_{l,l^\prime}+\alpha_{l}\alpha_{l'} e^{i \frac{2\pi n (l-l')}{N-1}} z_n(\nu_1-\nu_0) \\
 \nu_{N;l,l'}^{(1)} &=\delta_{l,l'}\left[\nu_0+\delta_{l,N-1}(\nu_1-\nu_0)\right],
 \end{aligned}
\end{equation}
where $\alpha_{l}=[N^{1/2}+\delta_{l,N-1}(1-N^{1/2})]/(N-1)$, and 
\begin{equation}\label{eq:phaseFactors}
 z_n=\exp\left[2\pi i\sum_{j=1}^n\left(\frac{1}{N}\sum_{k=1}^N q_k-q_j+\frac{Q}{N}\right)\right]
\end{equation}
is a phase factor (defined up to a global phase) that arises from Eq.~(\ref{eq:modifiedBoundaryCondition}). 
In Eq.~(\ref{eq:nu1}), the second term in the right hand side describes the enhancement of tunneling due to plasmon oscillations having a nonzero component along the direction of the phase slip. 
A simplified version of the procedure described above can be applied to recover~\cite{ivanov:01a} the ground-state tunneling amplitude, $\nu_n^{(0)}=z_n \nu_0$.

With no induced charges, the tunneling amplitudes are sensitive to the total charge $Q$.  
If $Q/N\in\mathbb{Z}$, we obtain [see Eqs.~(\ref{eq:splitting}), (\ref{eq:tunnelingAmplitudes}), (\ref{eq:nu1}), and (\ref{eq:phaseFactors})] $\nu^{(0)}=N\nu_0$ and
\begin{equation}\label{eq:defnEpsilon}
 \nu_{l,l'}^{(1)}=\delta_{l,l'} (N\nu_0+\epsilon)\,,\qquad \epsilon=(\nu_1-\nu_0)\frac{N}{N-1}\,,
\end{equation}
i.e., $\mu_n=(N\nu_0+\epsilon)^2$. 
Instead, if $Q/N\notin\mathbb{Z}$ it follows that $\nu^{(0)}=0$ ~\cite{ivanov:01a}, $\mu_1=0$ and $\mu_{n>1}=\epsilon^2$.
Thus, destructive interference caused by $Q/N\notin\mathbb{Z}$ protects a 2-fold degenerate crossing point.
The one-plasmon energy spectrum remains highly degenerate at any $Q/N$.

In reality, there are induced charges and there is no control over them~\cite{manucharyan:12a}. 
Below we regard $q_i$ as independent and identically distributed (i.i.d.)\ random variables,  and model them by Gaussian probability density functions (PDF) with mean zero and standard deviation $\sigma$. 
Consequently, $z_n$ and the tunneling amplitudes become random variables. 
For broad distributions ($\sigma\gtrsim 1$), the variables $z_n$ are approximately i.i.d.\ with uniformly distributed phase, and the tunneling amplitudes become insensitive to the value of $Q$.

At $N\gg 1$, the PDF for the ground state tunneling amplitude follows from the central limit theorem, 
\begin{equation}\label{eq:pdfZeroPlasmons}
\mathcal{P}(|\nu^{(0)}|)=\frac{2 |\nu^{(0)}|}{N{\nu_0}^2}e^{-\frac{|\nu^{(0)}|^2+{\nu_0}^2}{N {\nu_0}^2}}I_0\left(\frac{2|\nu^{(0)}|}{N\nu_0}\right)\hspace{-2pt},
\end{equation}
where $I_n(x)$ is the modified Bessel function of the first kind. 
From Eq.~(\ref{eq:pdfZeroPlasmons}), we obtain (to leading order in $N$) the expectation value $\langle |\nu^{(0)}|\rangle\simeq \nu_0 \sqrt{N}\sqrt{\pi}/2$. 

For the plasmon-assisted tunneling at $N\gg 1$ we find~\cite{supmat} $\mu_n=|\nu^{(0)}|^2+\epsilon^2+\eta_n$, where $\eta_n\approx 2 \epsilon {\rm Re}(\nu^{(0)} z_n^*)$.
Thus the conditional PDF of $|\mu_n|^{1/2}$ for a given $|\nu^{(0)}|$,
\begin{equation}\label{eq:conditionPDFofMu}
 \mathcal{P}\bigl(|\mu_n|^{1/2}\bigl||\nu^{(0)}|\bigr)\!=\frac{2}{\pi}\frac{|\mu_n|^{1/2}}{\sqrt{4\epsilon^2|\nu^{(0)}|^2-(|\mu_n|-|\nu^{(0)}|^2-\epsilon^2)^2}},
\end{equation}
 is finite in the window $\bigl||\nu^{(0)}|-\epsilon\bigr| < |\mu_n|^{1/2}< |\nu^{(0)}|+\epsilon$, see Fig.~\ref{fig:probDensFctOnePlasmon}. 
When $|\nu^{(0)}|>\epsilon$, the distribution of Eq.~(\ref{eq:conditionPDFofMu}) is approximately valid for all eigenvalues $\mu_n$.
When $\epsilon>|\nu^{(0)}|$, there is one eigenvalue that does not conform to Eq.~(\ref{eq:conditionPDFofMu}).
This single eigenvalue is smaller than $|\nu^{(0)}-\epsilon|$, and corresponds to a spatially delocalized mode~\cite{supmat,estimate}.   

The preceding paragraph evidences that the $2(N-1)$-fold degeneracy in the one-plasmon subspace of ${\cal H}_0$ is completely removed by phase slips in presence of random charges (see Fig.~\ref{fig:spectrum}). 
This is a manifestation of the Aharonov-Casher effect for plasmons. 
Inverting the unitary transformation that approximately diagonalizes $\nu^{(1)}\null^\dagger\nu^{(1)}$ (which turns out to diagonalize $\nu^{(1)}$ as well), and applying it onto the plasmon states $\psi_l$, 
we obtain $N-2$ modes ($N-1$, if $|\nu^{(0)}|>\epsilon$) that are almost perfectly localized in space.
Hence, in presence of randomly induced charges, phase slips lead to the localization of all-but-one (all, if $|\nu^{(0)}|>\epsilon$) plasmon excitations in single junctions. 
Without phase slips, the 
plasmon modes would be insensitive to induced charges.

Thus far we have assumed dispersionless plasmons, $C_0\simeq 0$ ~\cite{noteInducedCharges}.
For $C_0\neq 0$, the results derived above still hold provided that the bandwith of the plasmon dispersion ($\omega_p-\Omega_{l=1}$) is small compared to $\epsilon$.
In the opposite case, $N^2 C_0/C\gtrsim 16 \pi^2\epsilon/(\hbar\omega_p)$, the effect of $C_0$ on the plasmon energies can no longer be neglected. 
For large enough $C_0$, the typical energy separation between neighboring plasmon modes exceeds $\epsilon$, and the hybridization due to phase slips is restricted to subspaces of four-fold degenerate modes (two PCS and two plasmon modes $l$ and $N-l$). 
As long as $N^2C_0/C\ll 4 \pi^2$, the tunneling amplitudes between these four levels can be found by the method developed above, but with a set of plasmon modes diagonalizing Eq.~(\ref{eq:unperturbedHamiltonian}) for $C_0\neq 0$. 
In this case too, tunneling in presence of random induced charges removes all degeneracies. 
Yet, unlike for $C_0=0$, plasmons remain delocalized, thus weakening the Aharonov-Casher effect.

In summary, we have presented a theory for quantum phase slips in a closed Josephson junction array in a one-plasmon excited state.
Phase slips hybridize different one-plasmon and persistent current states, and lead to a typically non-degenerate energy spectrum that is sensitive to gate-induced charges.
This sensitivity is a signature of the Aharonov-Casher effect. 
Future work will address higher excited plasmon states, the interconvertion between current and plasmon excitations, and the effect of thermally excited quasiparticles.

We thank G. Blatter, M. Devoret, S. Nigg, and I. Pop for stimulating discussions. R.S. is grateful to Yale University for its hospitality throughout the duration of this project. 
I.G. and L.G. acknowledge the financial support from Yale University and DOE contract DE-FG02-08ER46482.





\appendix
\section{SUPPLEMENTAL MATERIAL}

This supplemental material has two parts. 
First, we provide technical details for the derivation of Eq.~(10) and Eq.~(11) in the main text.
Second, we provide technical details that justify Eq.~(16) in the main text,  and identify a single eigenvalue that can depart from it.
{\em Note about references:} the references cited below are listed at the end of the supplemental material.
\maketitle
\subsection{Coordinate transformations}
In this section we provide explicit expressions for the coordinate transformations that result in a decoupled effective Hamiltonian for the junction undergoing a phase slip and a quadratic Hamiltonian for the rest. 

The transformation of $\mathcal{H}$ [Eq.~(1)] to $\mathcal{H}_{\text{eff}}+\mathcal{H}_{\text{harm}}$ [Eq.~(10) and Eq.~(11)] to solve the first two of the three difficulties explained in the main text is most easily obtained via the Lagrangian $\mathcal{L}$. 
For $C_0=0$, a Legendre transformation yields
\begin{equation}\label{eq:lagrangian}
 \mathcal{L}=\frac{\hbar^2 C}{8e^2}\sum_{n=0}^{N-1}\dot\theta_{n}^2-E_J\sum_{n=0}^{N-1}\left[1-\cos\left(\theta_n+\frac{\delta}{N}\right)\right].
\end{equation}
Note that we labeled the variable $\theta_N$ by $\theta_0$ for convenience. 
Because only $N-1$ of the $N$ phase difference variables $\theta_i$ are independent [cf. Eq.~(2)], the variable $\theta_{N-1}$ can be replaced by
\begin{equation}\label{eq:replaceVariable}
 \theta_{N-1}=- \sum_{n=0}^{N-2} \theta_n \bmod{2\pi} \,.
\end{equation}
The phase differences enter Eq.~(\ref{eq:lagrangian}) only modulo $2\pi$. 
Therefore, Eq.~(\ref{eq:replaceVariable}) leads to
\begin{equation}\label{eq:lagrangian2}
\begin{aligned}
 \mathcal{L}&=\frac{\hbar^2 C}{8e^2}\sum_{n=0}^{N-2}\dot\theta_{n}^2-E_J\sum_{n=0}^{N-2}\left[1-\cos\left(\theta_n+\frac{\delta}{N}\right)\right] \\
  &+\frac{\hbar^2 C}{8e^2}\left(\sum_{n=0}^{N-2}\dot\theta_{n}\right)^2-E_J\left[1-\cos\left(\sum_{n=0}^{N-2}\theta_n-\frac{\delta}{N}\right)\right].
\end{aligned}
\end{equation}

We now focus on the case $\Phi=\Phi_0/2$, where $\delta=\pi$ and the lowest two PCS are $|m=0\rangle$ and $|m=-1\rangle$. 
As mentioned in the main text, we represent the $|m=0\rangle$ PCS by the minimum $\{\theta_i=0\}$. 
Let us first assume that the phase slip bringing the system into the $|m=-1\rangle$ PCS happens in junction $n_0\in\{0,\dots,N-2\}$; the case $n_0=N-1$ will be discussed at the end. 
We begin with a rotation of the coordinate system $\{\theta_0,\dots,\theta_{N-2}\}\to \{\tilde{\theta}_0,\dots,\tilde{\theta}_{N-2}\}$, which relabels the junction $n_0$ as junction $0$:
\begin{equation}\label{eq:rotationCoordinates}
 \begin{aligned}
  \tilde\theta_0 &\equiv \theta_{n_0}\,, \\
  \tilde\theta_k &\equiv \theta_{k+n_0}\,,\quad k\in\{1,\dots,N-2-n_0\} \,, \\
  \tilde\theta_k &\equiv \theta_{k+n_0-N+1}\,,\quad k\in\{N-1-n_0,\dots,N-2\} \,.
 \end{aligned}
\end{equation}
Next, a new set of variables $\{\tilde{\theta}_0,\dots,\tilde{\theta}_{N-2}\}\to\{y_0,\dots,y_{N-2}\}$ is constructed by choosing $y_0$ along the phase slip direction and completing the change of variables by a Gram-Schmidt process:
\begin{equation}
 \begin{aligned}
  y_0 &\equiv \frac{1}{\sqrt{N^2-N-1}}\left[(N-1)\tilde\theta_0-\sum_{j=1}^{N-2}\tilde\theta_j\right], \\
  y_1 &\equiv \frac{\sqrt{N-2}}{\sqrt{N^2-N-1}}\left[\tilde\theta_0+\frac{N-1}{N-2}\sum_{j=1}^{N-2}\tilde\theta_j\right], \\
  y_n &\equiv \frac{\sqrt{N-n-1}}{\sqrt{N-n}}\,\tilde\theta_{n-1}-\frac{1}{\sqrt{N-n}\sqrt{N-n-1}}\sum_{j=n}^{N-2}\tilde\theta_j\,, 
 \end{aligned}
\end{equation}
for $n\in\{2,\dots,N-2\}$.
These variables have the desired property that the two current minima are related to each other by the shift of one coordinate ($y_0$) only, while all other coordinate axes are orthogonal to it:
\begin{equation}\label{eq:oneVariableCondition}
 \begin{aligned}
  \theta_n &=0\quad &&\Leftrightarrow \quad\, y_i=0 \\ 
  \theta_n &=2\pi\delta_{n,n_0}-\frac{2\pi}{N} &&\Leftrightarrow\quad  \begin{aligned} y_0 &=2\pi\sqrt{N^2-N-1}/N \\ y_i &=0\quad i\in\{1,\dots,N-2\} \end{aligned}.
 \end{aligned}
\end{equation}

After making a harmonic approximation of the potential energies in Eq.~(\ref{eq:lagrangian2}) for all $\theta_{n\neq n_0}$ (the cosine term must be kept for the variable $\theta_{n_0}=\tilde{\theta}_0$), the resulting Lagrangian contains terms that couple different $y_i$.
This coupling is undesirable because it makes it difficult to evaluate the tunneling amplitudes. 

\begin{figure}[tb]
 \centering 
 \includegraphics{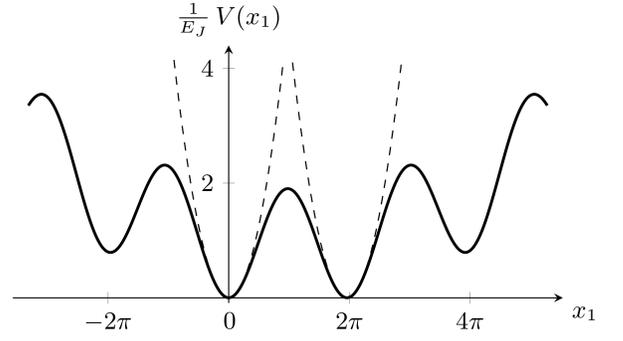}
 \caption{Effective potential energy seen by the junction undergoing a phase slip. Shown is the case $N=50$, together with the harmonic approximations (dashed lines) about $x_1=0$ (PCS $|m=0\rangle$) and $x_1=2\pi(N-1)/N$ (PCS $|m=-1\rangle$).}
 \label{fig:tunnelingPotential}
\end{figure}

The above problem may be resolved by making an additional linear transformation. 
A coordinate system that satisfies the idea of Eq.~(\ref{eq:oneVariableCondition}), while at the same time leading to a completely decoupled Hamiltonian with the cosine potential of junction $n_0$ preserved, can be found to be
\begin{equation}\label{eq:coordinateTrafoXVariables}
 \begin{aligned}
  x_1 &\equiv \frac{1}{\sqrt{N^2-N-1}}\left[(N-1)y_0+\sqrt{N-2}\,y_1\right] =\tilde{\theta}_0\,, \\
  x_2 &\equiv \frac{\sqrt{N^2-N-1}}{\sqrt{N-1}}\,y_1=\frac{\sqrt{N-2}}{\sqrt{N-1}}\,\tilde\theta_0+\frac{\sqrt{N-1}}{\sqrt{N-2}}\sum_{n=1}^{N-2}\tilde\theta_n \,, \\
  x_n &\equiv y_{n-1}\,,\qquad n\in\{3,\dots,N-1\}\,.
 \end{aligned}
\end{equation}
After this transformation, Eq.~(\ref{eq:oneVariableCondition}) changes to
\begin{equation}\label{eq:oneVariableConditionX}
 \begin{aligned}
  \theta_n &=0  &&\Leftrightarrow\quad \,x_1=0\quad \\ 
  \theta_n &=2\pi\delta_{n,n_0}-\frac{2\pi}{N} &&\Leftrightarrow\quad \begin{aligned} x_1 &=2\pi(N-1)/N \\  x_i &=0\quad i\in\{2,\dots,N-1\}\end{aligned}
 \end{aligned}
\end{equation}
and the Lagrangian reads
\begin{equation}\label{eq:effectiveLagrangian}
 \mathcal{L} =\frac{1}{2}M\frac{N}{N-1}\, \dot{x}_1^2-V(x_1)+\frac{1}{2}M\sum_{n=2}^{N-1}\left(\dot{x}_n^2-\omega_p^2 x_n^2\right)\,,
\end{equation}
where (cf. Fig.~\ref{fig:tunnelingPotential})
\begin{equation}\label{eq:tunnelingPotential}
 V(x_1)\equiv E_J \left[1-\cos\left(x_1+\frac{\pi}{N}\hspace{-1pt}\right)-\frac{\pi}{N}\,x_1+\frac{1}{2(N-1)}\, x_1^2\right]
\end{equation}
is the effective potential seen by the junction undergoing the phase slip. 
In Eq.~(\ref{eq:effectiveLagrangian}), $M=\hbar^2/(8E_C)$ and $\omega_p$ is defined in the main text [Eq.~(6)]. 
Note that Eq.~(\ref{eq:effectiveLagrangian}) contains no terms that couple different $x_i$.
From there, $\mathcal{H}\to\mathcal{H}_{\text{eff}}+\mathcal{H}_{\text{harm}}$ follows immediately, with Eq.~(10) and Eq.~(11).

For completeness, we provide the inverse transformation of Eq.~(\ref{eq:coordinateTrafoXVariables}), which is necessary to arrive at the results presented in the main text:
\begin{equation}
 \begin{aligned}
  \tilde{\theta}_0 &=x_1\,, \\ 
  \tilde{\theta}_n &=  -\frac{1}{N-1}\, x_1+\frac{1}{\sqrt{N-2}\sqrt{N-1}}\, x_2 \\ 
   &	-\sum_{j=2}^{n}\frac{1}{\sqrt{N-j}\sqrt{N-j-1}}\,x_{j+1}+ \frac{\sqrt{N-n-2}}{\sqrt{N-n-1}}\, x_{n+2}
 \end{aligned}
\end{equation}
for $n\in\{1,\dots,N-2\}$.
In addition, we note the useful identity
\begin{equation}\label{eq:usefulIdentity}
 \sum_{n=0}^{N-2}\tilde{\theta}_n+\Biggl(\sum_{j=0}^{N-2}\tilde{\theta}_j\Biggr)^2=\frac{N}{N-1}\, x_1^2+\sum_{n=2}^{N-1}x_n^2\,.
\end{equation}

We now discuss the case $n_0=N-1$. 
The above changes of variables do not work for this case. 
Nevertheless, following the same ideas still leads to a suitable set of new variables. 
It turns out that real and imaginary parts of the discrete Fourier transform of $\{\theta_n\}$, 
\begin{equation}\label{eq:defnOfZeta}
 \tilde{x}_{k+1} \equiv
 \begin{cases} -\sqrt{N-1}\re(\xi_k) & k=0 \\ \sqrt{2} \re(\xi_k) & k\in\{1,\dots,{N}_h-1\} \\ \sqrt{2} \re(\xi_k) & k={N}_h\quad\text{$N$ even} \\ \re(\xi_k) & k={N}_h\quad\text{$N$ odd} \\ \sqrt{2}\im(\xi_{k-{N}_h}) & k\in\{{N}_h+1,\dots,N-2\} \end{cases}
\end{equation}
with
\begin{equation}
 {N}_h\equiv\begin{cases} \frac{N-1}{2} & N \text{ odd} \\ \frac{N-2}{2} & N\text{ even} \end{cases},
\end{equation}
and 
\begin{equation}\label{eq:defnOfXi}
 \xi_k \equiv \frac{1}{\sqrt{N-1}}\sum_{n=0}^{N-2}\theta_n e^{i\frac{2\pi}{N-1}kn}\,,\quad k\in\{0,\dots,N-2\}\,,
\end{equation}
lead (upon replacing $x_n$ by $\tilde{x}_n$) to exactly the same Lagrangian [Eq.~(\ref{eq:effectiveLagrangian})], and with it to the same Hamiltonian $\mathcal{H}_{\text{eff}}+\mathcal{H}_{\text{harm}}$, while respecting Eq.~(\ref{eq:oneVariableConditionX}).
For these variables the inverse transformation is essentially given by
\begin{equation}
 \theta_n = \frac{1}{\sqrt{N-1}}\sum_{k=0}^{N-2}\xi_k e^{-i\frac{2\pi}{N-1}kn}\,,
\end{equation}
and Eq.~(\ref{eq:usefulIdentity}) still holds.
Even though the contribution of junction $N-1$ might seem different at first, it leads to exactly the same effective Hamiltonian as the other junctions, as expected from symmetry arguments.

\subsection{Energy-splittings in presence of random charges}

In this section we derive the distribution of the energy splittings in the one-plasmon subspace, and discuss the existence of a ``zero mode'' in the limit $\epsilon\gg|\nu^{(0)}|$.

The energy-splittings are given by $2|\mu_n|^{1/2}$, where $\mu_n$ ($n\in\{1,\dots,N-1\}$) are the eigenvalues of the hermitian matrix $M\equiv (\nu^{(1)})^\dagger\nu^{(1)}$.
Using Eq.~(12), it is straightforward to obtain $M_{l l'}$ in the eigenbasis of plasmon modes; however, the resulting expression is cumbersome and will not be shown here.
The diagonalization of $M$ becomes significantly simpler after Fourier transforming to (junction) coordinate space:
\begin{equation}
M_{n n'}=\frac{1}{N-1} \sum_{l,l'=1}^{N-1} e^{-i\frac{2\pi n l}{N-1}} e^{i\frac{ 2\pi n' l'}{N-1}} M_{l l'}.
\end{equation}
This results in $M_{n n'}=M_{n n'}^{(0)}+M_{n n'}^{(1)}$, where
\begin{align}
\label{eq:M}
M_{n n'}^{(0)} &=\left[|\nu^{(0)}|^2+\epsilon^2+2\,\epsilon\, {\rm Re}(\nu^{(0)} z_n^*)\right]\delta_{n,n'}\nonumber\\
M_{n n'}^{(1)} &=\frac{2\epsilon}{N}{\rm Re}\left[\nu^{(0)}(1-z_n^*-z_{n'}^*)\right]-\frac{\epsilon^2}{N} z_n^* z_{n'},
\end{align}
and we have neglected subleading terms in $1/N$. 
Since the matrix elements of $M^{(1)}$ are $N$ times smaller than those of $M^{(0)}$, we can at first approximate $M_{n n'}\simeq M^{(0)}_{n n'}$.
This yields
\begin{equation}
\label{eq:mu0}
\mu_n^{(0)}= |\nu^{(0)}|^2+\epsilon^2+ 2\,\epsilon\,{\rm Re}(\nu^{(0)} z_n^*),
\end{equation}
which is the result quoted above Eq.~(16) in the main text.
After noticing that $|\nu^{(0)}|$ and $z_n$ are approximately independent distributed for $N\gg 1$, we plug Eq.~(\ref{eq:mu0}) in the standard definition of the conditional probability~\cite{reichl} and straightforwardly arrive at Eq.~(16).
The corresponding eigenfunctions are spatially localized, because (i) $M_{n n'}^{(0)} \propto \delta_{n,n'}$, and (ii) $\mu_n^{(0)}$ are typically non-degenerate (due to random charges).
This implies that plasmon modes get spatially localized (at $C_0=0$) in presence of phase slips and random charges.

So far we have neglected $M^{(1)}$. 
Although its matrix elements are $N$ times smaller than those of $M^{(0)}$, there are $N$ times as many of them.
This suggests that the effect of $M^{(1)}$ on $\mu_n$ can be relevant, at least for some of the eigenvalues. 
Here we will analyze this issue analytically in the regimes $\epsilon\gg|\nu^{(0)}|$ and $\epsilon\ll|\nu^{(0)}|$.
We will argue that Eq.~(16) remains a good approximation for (at most) all-but-one eigenvalues.
This statement is in agreement with what we find from a numerical diagonalization of $(\nu^{(1)})^\dagger\nu^{(1)}$ for arbitrary $|\nu^{(0)}|/\epsilon$.

Let us begin from the case $\epsilon\gg|\nu^{(0)}|$, which is relevant for some experiments (e.g. $\epsilon\simeq 10 \langle|\nu^{(0)}|\rangle$ in Ref.~\cite{manu}).
In this regime, we may approximate
\begin{equation}
M_{n n'}^{(1)} \simeq -\frac{\epsilon^2}{N} z_n^* z_{n'}.
\end{equation}
Due to the separability of $M_{n n'}^{(1)}$, $M^{(0)}_{n n'}+M_{n n'}^{(1)}$ can be diagonalized analytically.
The eigenvalue equation reads 
\begin{equation}
\sum_{n'} \left[M^{(0)}_{n n'}+M^{(1)}_{n n'}\right]\psi^{(\lambda)}_{n'}=\mu_\lambda \psi^{(\lambda)}_{n},
\end{equation}
which can be rewritten as
\begin{equation}
\label{eq:psi}
\psi^{(\lambda)}_n=\frac{\epsilon^2}{N} \frac{ z_n^* c_\lambda}{M_{n n}^{(0)}-\mu_\lambda},
\end{equation}
where $\lambda$ is the eigenvalue label and
\begin{equation}
\label{eq:C}
c_\lambda\equiv\sum_{n'} z_{n'}\psi^{(\lambda)}_{n'}.
\end{equation}
Substituting Eq.~(\ref{eq:psi}) in Eq.~(\ref{eq:C}) and using $|z_n|^2=1$, we get
\begin{equation}
c_\lambda=c_\lambda\frac{\epsilon^2}{N}\sum_n\frac{1}{M_{n n}^{(0)}-\mu_\lambda}.
\end{equation}
If $c_\lambda\neq 0$, this gives
\begin{equation}
\label{eq:gra}
1=\frac{\epsilon^2}{N}\sum_{n=1}^{N-1}\frac{1}{|\nu^{(0)}|^2+\epsilon^2+2\,\epsilon\, {\rm Re}(\nu^{(0)} z_n^*)-\mu_\lambda}.
\end{equation}
The right hand side of this equation is a sum of $N-1$ terms with poles at $\mu_\lambda=\mu_\lambda^{(0)}$.
For a given configuration of $z_n$ and for $N\gg 1$, a graphical solution of Eq.~(\ref{eq:gra}) shows that $N-2$ eigenvalues, given by $\mu_\lambda\simeq\mu_\lambda^{(0)}$ ($\lambda\in\{2,\dots,N-1\}$), are densely packed between $(\epsilon-|\nu^{(0)}|)^2$ and $(\epsilon+|\nu^{(0)}|)^2$.
Their respective eigenmodes coincide with those of $M^{(0)}$, i.e. they are spatially localized.
In summary, Eq.~(16) remains valid for $N-2$ eigenvalues.

However, the graphical representation also shows the existence of a single eigenvalue $\mu_1$ such that $\mu_1<(\epsilon-|\nu^{(0)}|)^2$.
The appearance of this eigenvalue outside the dense distribution formed by all the other eigenvalues is mathematically analogous to what happens in the Cooper problem of superconductivity~\cite{schrieffer}.
There, the lowest eigenvalue (equal to the binding energy of a Cooper pair) lies below a quasicontinuum energy spectrum for pairs of unbound electrons.

Anticipating that $\mu_1\ll|\nu^{(0)}|^2+\epsilon^2-2\epsilon|\nu^{(0)}|$, we rewrite Eq.~(\ref{eq:gra}) as
\begin{equation}
1\simeq\frac{\epsilon^2}{N}\sum_n\frac{1}{|\nu^{(0)}|^2+\epsilon^2-\mu_1}\left(1-\frac{2\,\epsilon\, {\rm Re}(\nu^{(0)} z_n^*)}{|\nu^{(0)}|^2+\epsilon^2-\mu_1}\right).
\end{equation}
A simple calculation then shows that
\begin{equation}
\mu_1\simeq |\nu^{(0)}|^2+\frac{2\epsilon|\nu^{(0)}|^2}{N \nu_0},
\end{equation}
where we have used $\nu^{(0)}=\nu_0\sum_n z_n$.
Thus, $2 |\mu_1|^{1/2}$ agrees approximately with the ground-state energy-splitting.  
Since $\epsilon\gg|\nu^{(0)}|$, $\mu_1\ll\mu_{\lambda>1}$ and thus the $\lambda=1$ eigenvalue may be loosely referred to as a ``zero mode''.
Its eigenfunction is
\begin{equation}
\psi_n^{(\lambda=1)}\propto \frac{z_n^*}{M_{n n}^{(0)}-\mu_1}
\end{equation}
modulo a $n-$independent normalization factor.
For $\epsilon\gg |\nu^{(0)}|$, the weight $|\psi_n^{(1)}|^2$ is approximately independent of $n$ and hence the ``zero mode'' is completely delocalized onto all the junctions in the ring.
As $\epsilon/|\nu^{(0)}|$ becomes smaller, $|\mu_{\lambda\neq 1}-\mu_1|$ decreases, localization effects start to become visible in  $\psi_n^{(1)}$, and the term ``zero mode'' gradually becomes meaningless.

In the above derivation we have assumed $c_\lambda\neq 0$.
If $c_\lambda=0$, we are constrained to a $N-2$ -dimensional plane (orthogonal to the vector corresponding to $\lambda=1$) and as expected we get $\mu_\lambda=\mu_\lambda^{(0)}$.

Finally, let us briefly consider the opposite case $|\nu^{(0)}|\gg \epsilon$.
In this regime, we may approximate
\begin{equation}
M_{n n'}^{(1)} \simeq \frac{2 \epsilon}{N} {\rm Re}[\nu^{(0)}(1- z_n^*- z_{n'}^*)]
\end{equation}
and proceed to analyze the corresponding eigenvalue equation much like for the $\epsilon\gg|\nu^{(0)}|$ case.
This eigenvalue equation is less simple than the one above; it reads
\begin{equation}
\label{eq:mess}
-1=\frac{\int_{-1}^{1} dx\frac{\rho(x)}{x-1-\xi_\lambda} \int_{-1}^{1} dx\frac{\rho(x) x(1-x)}{x-1-\xi_\lambda}}{\left[1+\int_{-1}^{1} dx\frac{\rho(x) (1-x)}{x-1-\xi_\lambda}\right]\left[1-\int_{-1}^{1} dx\frac{\rho(x) x}{x-1-\xi_\lambda}\right]},
\end{equation}
where $\xi_\lambda\equiv (\mu_\lambda-\epsilon^2-|\nu^{(0)}|^2-2\epsilon|\nu^{(0)}|)/(2\epsilon|\nu^{(0)}|)$, and
\begin{equation}
\label{eq:rho}
\rho(x)=\frac{1}{N}\sum_{n=1}^{N-1} \delta(x-{\rm Re}(z_n)).
\end{equation}
The function $\rho(x)$ is random.
In the derivation of Eq.~(\ref{eq:mess}), we have made a phase rotation so that $\nu^{(0)}\to |\nu^{(0)}|$. 
We can gain some insight into the solution of Eq.~(\ref{eq:mess}) by replacing $\rho(x)\to\langle\rho(x)\rangle\simeq (1/\pi)(1-x^2)^{-1/2}$, upon which Eq.~(\ref{eq:mess}) can be solved numerically.
The solution shows that $\mu_\lambda\simeq \mu_\lambda^{(0)}$ is a good approximation for {\em all} $N-1$ eigenvalues (provided that $N\gg 1$). 
In other words, there is no eigenvalue that lies outside the distribution of Eq.~(16).
Thus, the appearance of a delocalized ``zero mode'' is exclusive to the regime $\epsilon\gg|\nu^{(0)}|$.


\begin{thebibliography}{99}
\bibitem{N}
 The $N$-th ``mode'' corresponds to a simultaneous shift of all $\varphi_i$, and can be ignored because we consider a fixed $Q$.
\bibitem{symmetric}
The symmetric notation is obtained by replacing $N-l$ by $-l$ for $l>N/2$.
\bibitem{nu0}
 $\nu_0$ and $\nu_1$ are evaluated using ${\cal H}_{\rm eff}$. 
At large $N$, the calculation consists of solving the Mathieu equation~\cite{mathieu}. 
For $E_J\gg E_C$, one finds $\nu_0\simeq (4/\sqrt{\pi}) (8 E_J^3 E_C)^{1/4} \exp[-(8 E_J/E_C)^{1/2}]$ and $\nu_1/\nu_0\simeq 4 (8 E_J/E_C)^{1/2}$.
\bibitem{estimate}
We note that for the parameters of experiment~\cite{manucharyan:12a}, $E_J=22\, {\rm GHz}$, $E_C=1\, {\rm GHz}$, $N=43$, one finds $\langle|\nu^{(0)}|\rangle\simeq 0.4\, {\rm M Hz}\ll \epsilon\simeq 3.5\, {\rm M Hz}$.
\bibitem{noteInducedCharges}
 We assumed small $C_0$ but allowed for large enough $V_i$, such that $q_i=C_0V_i/2e$ could be finite.
\end{thebibliography}
\end{document}